\documentstyle[prl,aps,amsfonts,amssymb,epsf,floats]{revtex}
\begin{document}
\draft
\twocolumn[\hsize\textwidth\columnwidth\hsize\csname@twocolumnfalse\endcsname

\title{
Effects of point defects on the phase diagram of vortex states\\
in high-$T_{\rm c}$ superconductors in $\vec{B} \parallel c$ axis
}

\author{Yoshihiko Nonomura \cite{nonomail} and Xiao Hu}
\address{
National Research Institute for Metals, 
Tsukuba, Ibaraki 305-0047, Japan}
\date{November 17, 2000}
\maketitle
\begin{abstract}
The phase diagram for the vortex states of high-$T_{\rm c}$ 
superconductors with point defects in $\vec{B} \parallel c$ axis 
is drawn by large-scale Monte Carlo simulations. The vortex 
slush (VS) phase is found between the vortex glass (VG) and 
vortex liquid (VL) phases. The first-order transition between 
this novel normal phase and the VL phase is characterized by 
a sharp jump of the density of dislocations. The first-order 
transition between the Bragg glass (BG) and VG or VS phases 
is also clarified. These two transitions are compared with 
the melting transition between the BG and VL phases. 
\end{abstract}
]
%
Vortex states in high-$T_{\rm c}$ superconductors in 
$\vec{B}\parallel c$ axis have been intensively studied. 
Although the melting transition in pure systems has now 
been understood very well, experimental phase diagrams 
are more complicated owing to effects of impurities. 
In the present Letter, point defects are taken into 
account as the simplest impurity. 

\begin{figure}
\epsfxsize = 7.0cm
\epsffile{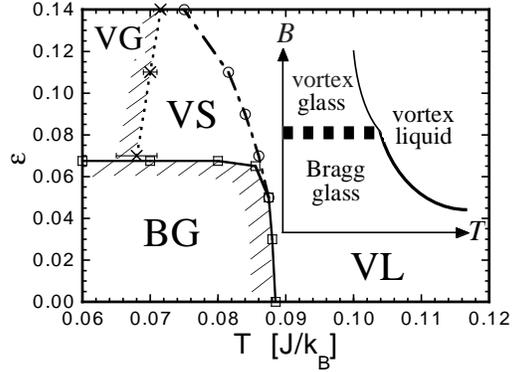}
\vspace{0.2cm}
\caption{
$\epsilon$--$T$ phase diagram of the model with 
$\Gamma=20$. Shaded phases exhibit superconductivity. 
Schematic $B$--$T$ phase diagram is shown in the inset 
for comparison. 
}
\label{phasefig}
\end{figure}

A schematic phase diagram of high-$T_{\rm c}$ vortex states 
with point defects is given in the inset of Fig.\ \ref{phasefig}. 
These three phases have been observed in 
YBa$_{2}$Cu$_{3}$O$_{7-\delta}$ (YBCO) \cite{Safar95,Nishizaki98} and 
Bi$_{2}$Sr$_{2}$CaCu$_{2}$O$_{8+y}$ (BSCCO). \cite{Cubitt93,Khaykovich97}
The Bragg glass (BG) phase \cite{Giamarchi} is characterized 
by the power-law decay of correlation functions of vortex 
positions \cite{Nattermann,Giamarchi} and the triangular Bragg 
pattern of the structure factor. The vortex glass (VG) phase 
was first defined on the basis of phase variables, \cite{Fisher} 
and can be defined alternatively on the basis of vortex 
positions. \cite{Nattermann00} Recent simulations including 
screening effects \cite{Pfeiffer,Kawamura} suggest the 
absence of the phase-coherent VG. However, the positional 
VG is expected to be more stable than the phase-coherent 
VG. \cite{Nattermann00} Therefore, whether the VG exists 
as the thermodynamic phase or not is still unsettled. 

The BG--VG phase boundary was studied 
analytically, \cite{Ertas96,Kierfeld97,Giamarchi97} 
essentially based on the Lindemann criterion. 
Numerically, difference between the BG and VG phases was 
discussed, \cite{Gingras,Ryu,Olson} and the overall phase 
diagram was obtained recently. \cite{Otterlo,Sugano} However, 
studies based on thermodynamic quantities are still lacking. 

Quite recently, another phase has been observed 
experimentally. Nishizaki {\it et al.} have found a sharp 
jump of the resistivity \cite{Nishizaki00,Nishizaki01} and the 
magnetization \cite{Nishizaki01} in optimally-doped YBCO 
in the vortex liquid (VL) region. They pointed out that 
this might be the transition to the vortex slush (VS) 
phase \cite{Worthington} originally found in irradiated YBCO. 
The VS phase is characterized by evolution of a short-range 
order, and the boundary between the VS and VL phases 
terminates at a critical point. Similar anomalies were also 
observed in BSCCO. \cite{Fuchs,Blasius,Khaykovich00,Kadowaki00} 

In the present Letter, we compose the phase diagram 
in the pinning-strength ($\epsilon$)--temperature 
($T$) plane (Fig.\ \ref{phasefig}) on the basis of 
thermodynamic quantities. 
The first-order melting transition occurs between the 
BG and VL phases as in pure systems. \cite{Hu,Nono99} 
A first-order transition line stretches from the melting 
line into the VL region dividing the VS and VL phases.  
The VG phase exists at much lower temperatures. 
The boundary between the BG and VG or VS phases is almost 
independent of temperature, and the phase transition is 
of first order. This phase diagram is quite similar to 
the $B$--$T$ phase diagram observed experimentally. 

Our formulation is based on the three-dimensional frustrated 
XY model on a simple cubic lattice, \cite{Li,Hu}
\begin{eqnarray}
  \label{XYham}
  \cal{H}&=&-\hspace{-0.2cm}
                 \sum_{i,j \in ab\ {\rm plane}} \hspace{-0.3cm}
                 J_{ij}\cos \left(\phi_{i}-\phi_{j}-A_{ij}\right)
                 \nonumber\\
           &&-\frac{J}{\Gamma^{2}} \hspace{-0.1cm}
               \sum_{m,n\parallel c\ {\rm axis}}\hspace{-0.2cm}
               \cos \left(\phi_{m}-\phi_{n}\right)\ ,\\
  A_{ij}&=&\frac{2\pi}{\Phi_{0}}\int^{j}_{i}{\bf A}^{(2)}
                                      \cdot {\rm d}{\bf r}^{(2)},
  \ \ \vec{B}=\vec{\nabla}\times \vec{A}\ ,
\end{eqnarray}
with the periodic boundary condition. Screening effects are not 
included in this model. A uniform magnetic field $\vec{B}$ is 
applied along the $c$ axis, and the averaged number of flux lines 
per plaquette is given by $f=l^{2}B/\Phi_{0}$. Here $l$ stands for 
the grid spacing in the $ab$ plane. Point defects are introduced as 
the plaquettes which consist of four weaker couplings and are 
randomly distributed in the $ab$ plane with probability $p$. 
Couplings are given by $J_{ij}=(1-\epsilon)J$ ($0<\epsilon<1$) on 
the point defects, and $J_{ij}=J$ elsewhere. Here we concentrate on 
the model with $\Gamma=20$, $f=1/25$ and $p=0.003$. We do not 
consider the lower critical point \cite{Roulin,Paulius} at present. 
The system size is $L_{x}=L_{y}=50$ and $L_{c}=40$, which is 
large enough to analyze the pure system ($\epsilon=0$). 

For each $\epsilon$, Monte Carlo simulations are started from a 
very high temperature, and systems are gradually cooled down. 
After such annealing, further equilibration and measurement 
are performed at each temperature. Typical simulation times 
are $4\sim 12\times 10^{7}$ Monte Carlo steps (MCS) for 
equilibration, and $2\sim 5\times 10^{7}$ MCS for 
measurement. The present simulations are based on 
one sample. Since configurations of point defects are 
independent in different $ab$ planes, it is reasonable 
to expect that there exists a self-averaging effect. 
We calculate the internal energy $e$, the specific 
heat $C$, the helicity modulus along the $c$ axis 
$\Upsilon_{\rm c}$, \cite{Li,Hu} and the phase 
difference between the nearest-neighbor $ab$ 
planes $\langle \cos(\phi_{n}-\phi_{n+1})\rangle$, 
together with the ratio of entangled flux lines to total 
flux lines $N_{\rm ent}/N_{\rm flux}$, \cite{Nono99} the 
density of dislocations in the $ab$ plane $\rho_{\rm d}$, 
and the structure factor of flux lines in the $ab$ plane. 
The helicity modulus is proportional to the superfluid 
density, and is the order parameter of superconductivity. 
The inter-layer phase difference is related \cite{Bulaevskii} 
to the frequency of the Josephson plasma resonance 
(JPR).\ \cite{Tachiki94,Matsuda95} 

\begin{figure}
\epsfxsize = 6.5cm
\epsffile{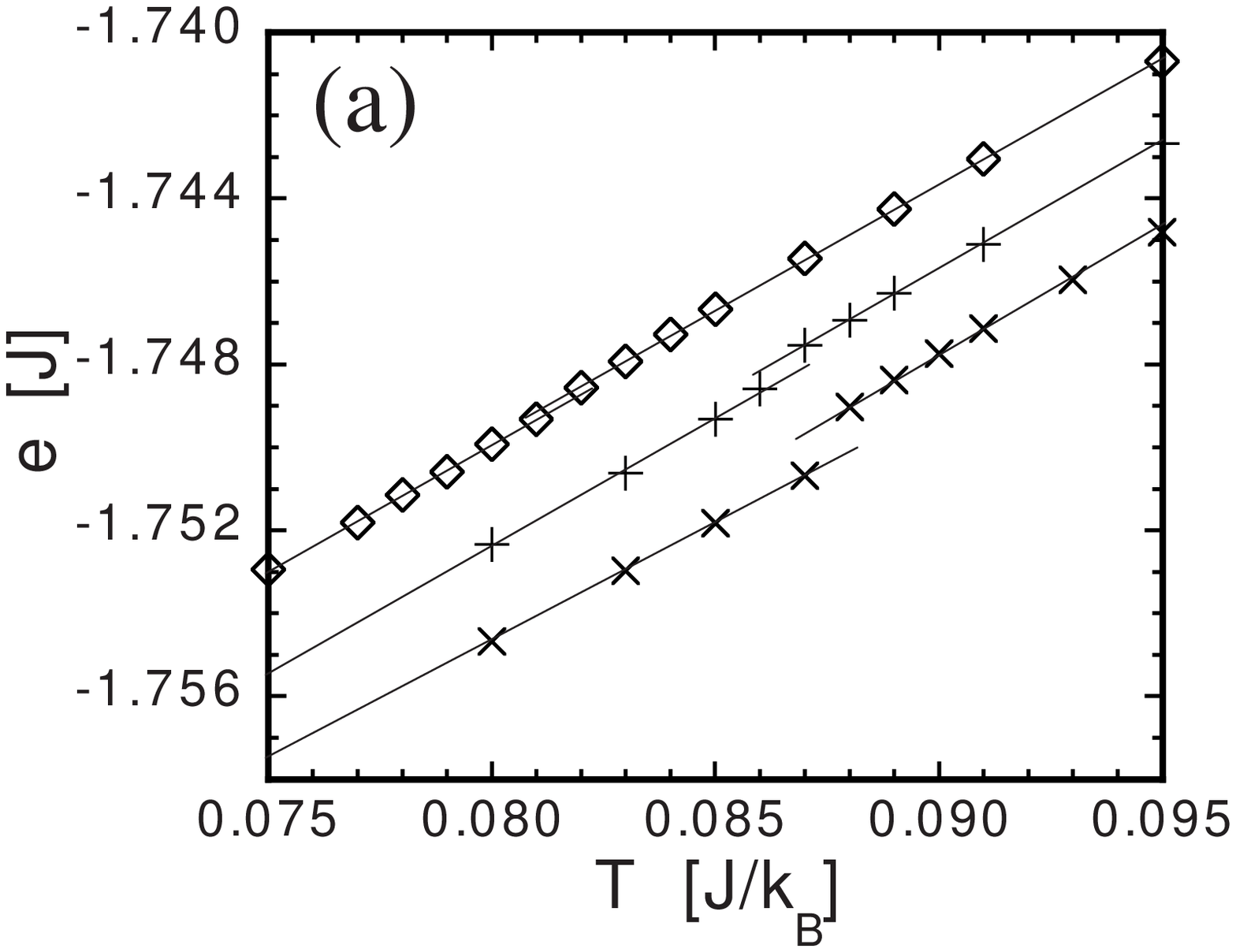}
\epsfxsize = 7.2cm
\epsffile{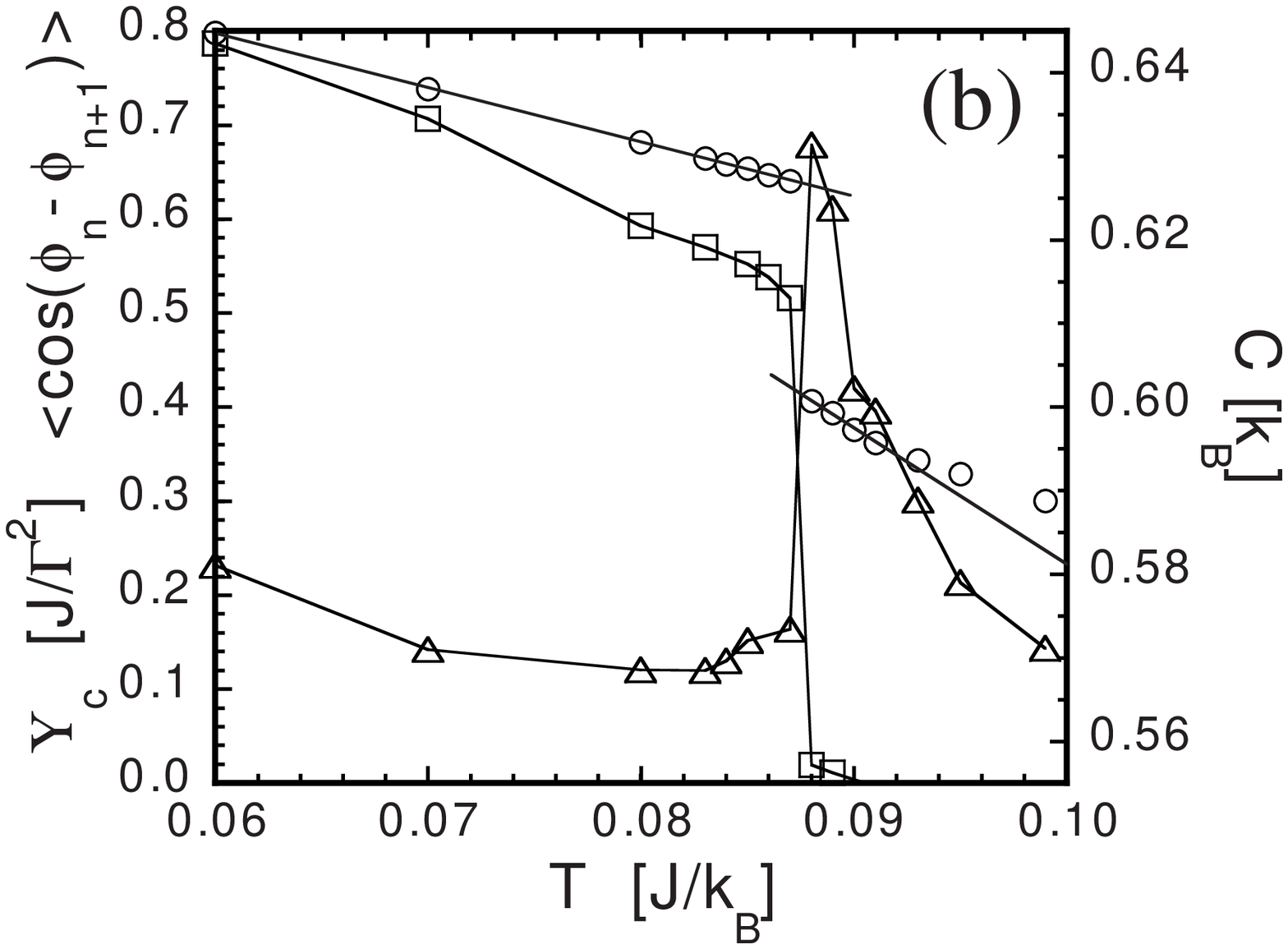}
\epsfxsize = 7.2cm
\epsffile{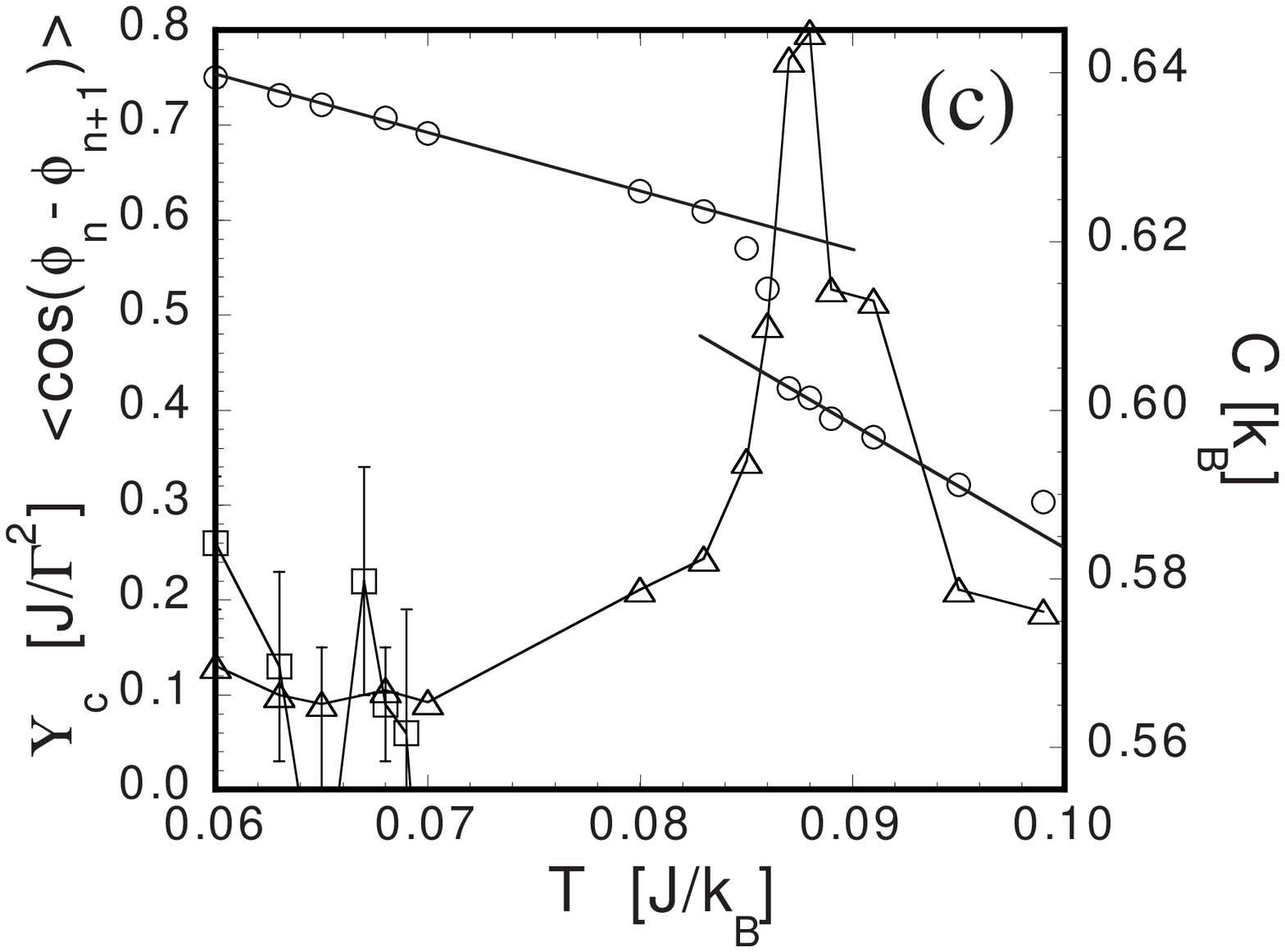}
\epsfxsize = 7.2cm
\epsffile{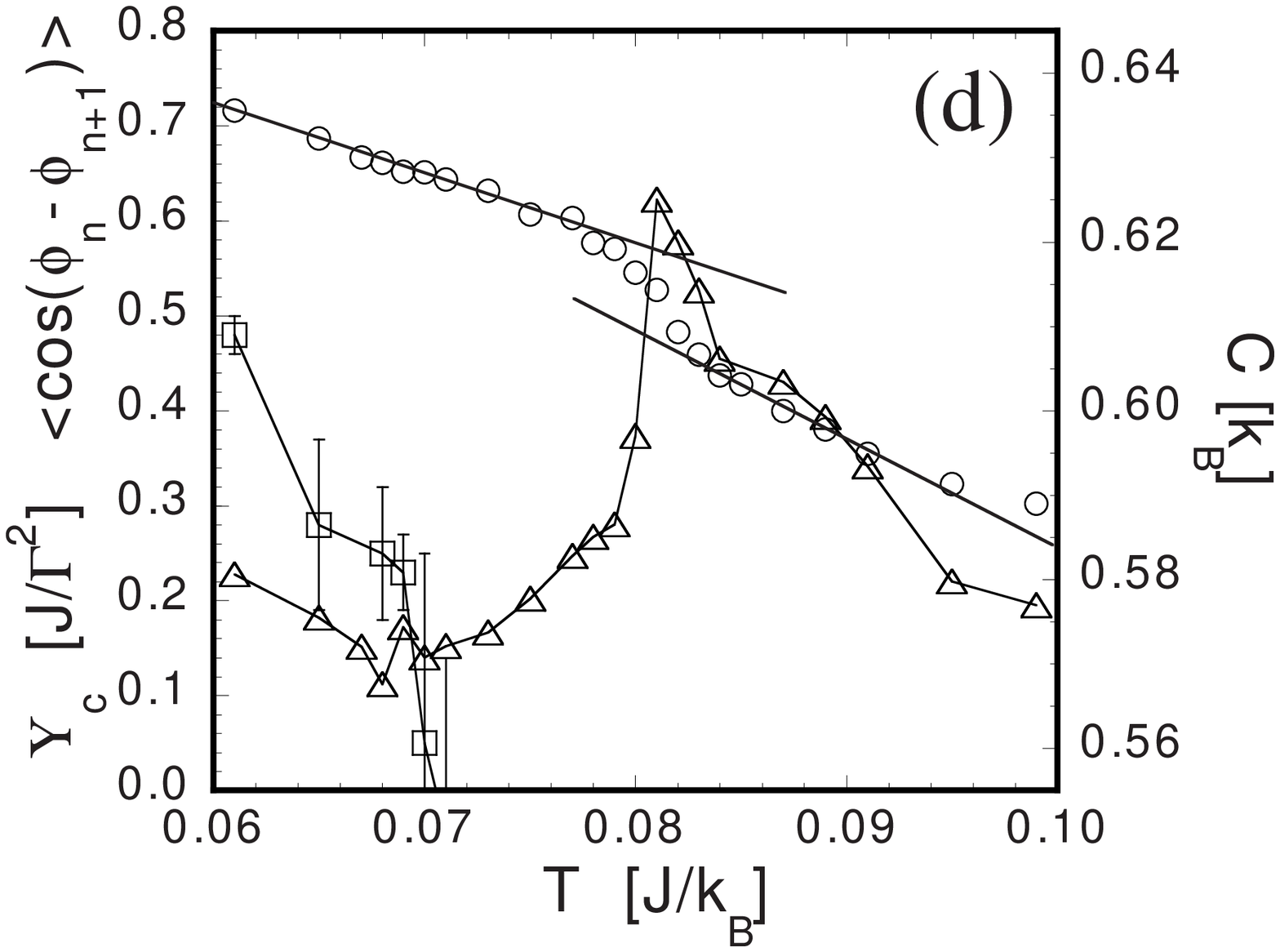}
\vspace{0.2cm}
\caption{
Temperature dependence of (a) $e$ at $\epsilon=0.05$ 
($\times$), $0.07$ ($+$) and $0.11$ ($\diamond$), and 
$C$ ($\triangle$), $\Upsilon_{\rm c}$ ($\Box$) and 
$\langle\cos(\phi_{n}-\phi_{n+1})\rangle$ ($\circ$) 
at (b) $\epsilon=0.05$, (c) $0.07$ and (d) $0.11$.
The origin of $e$ is varied for each $\epsilon$ in (a). 
}
\label{VSVLfig}
\end{figure}
{\it BG--VL and VS--VL phase transitions}.---Temperature dependence of 
$e$, $C$, $\Upsilon_{c}$ and $\langle\cos(\phi_{n}-\phi_{n+1})\rangle$ 
at $\epsilon=0.05$, $0.07$ and $0.11$ is displayed in Figs.\ 
\ref{VSVLfig}(a)--(d). The BG--VL transition is observed at 
$\epsilon=0.05$, and the VS--VL one at $\epsilon=0.07$ and $0.11$. 
Jumps of $e$ and $\langle\cos(\phi_{n}-\phi_{n+1})\rangle$ and the 
$\delta$-function peak of $C$ are observed at the melting temperature 
$T_{\rm m}$ or the slush temperature $T_{\rm sl}$. Finite latent heats 
$Q=\Delta e$ indicate that these phase transitions are of first order. 
The anomalies of the VS--VL transition at $\epsilon=0.07$ are as large 
as those of the BG--VL transition at $\epsilon=0.05$, and gradually 
lose sharpness as $\epsilon$ is increased (see Figs.\ \ref{VSVLfig}(c) 
and \ref{VSVLfig}(d)). At $\epsilon=0.14$, no anomalies are observed 
other than a small jump of $\langle\cos(\phi_{n}-\phi_{n+1})\rangle$. 
These properties can be explained by the existence of the critical 
point \cite{Nishizaki00,Nishizaki01,Worthington} which terminates 
the first-order VS--VL transition line around $\epsilon=0.14$. 

The most important difference between the BG--VL and VS--VL transitions is 
seen in $\Upsilon_{c}$. In the BG--VL transition, this quantity appears 
discontinuously at $T_{\rm m}$. On the other hand, $\Upsilon_{c}$ remains 
vanishing for $T<T_{\rm sl}$ in the VS--VL transition. In other words, 
the BG--VL phase transition is the superconducting--normal one, 
while the VS--VL transition occurs between two normal phases. 
The latter transition does not contradict the existence of the 
critical point. The proliferation of $\Upsilon_{c}$ at lower 
temperatures signals the phase transition to the VG phase, 
though error bars of $\Upsilon_{c}$ are fairly large 
owing to very large correlation time in the VG phase. 

\begin{figure}
\epsfxsize = 7.2cm
\epsffile{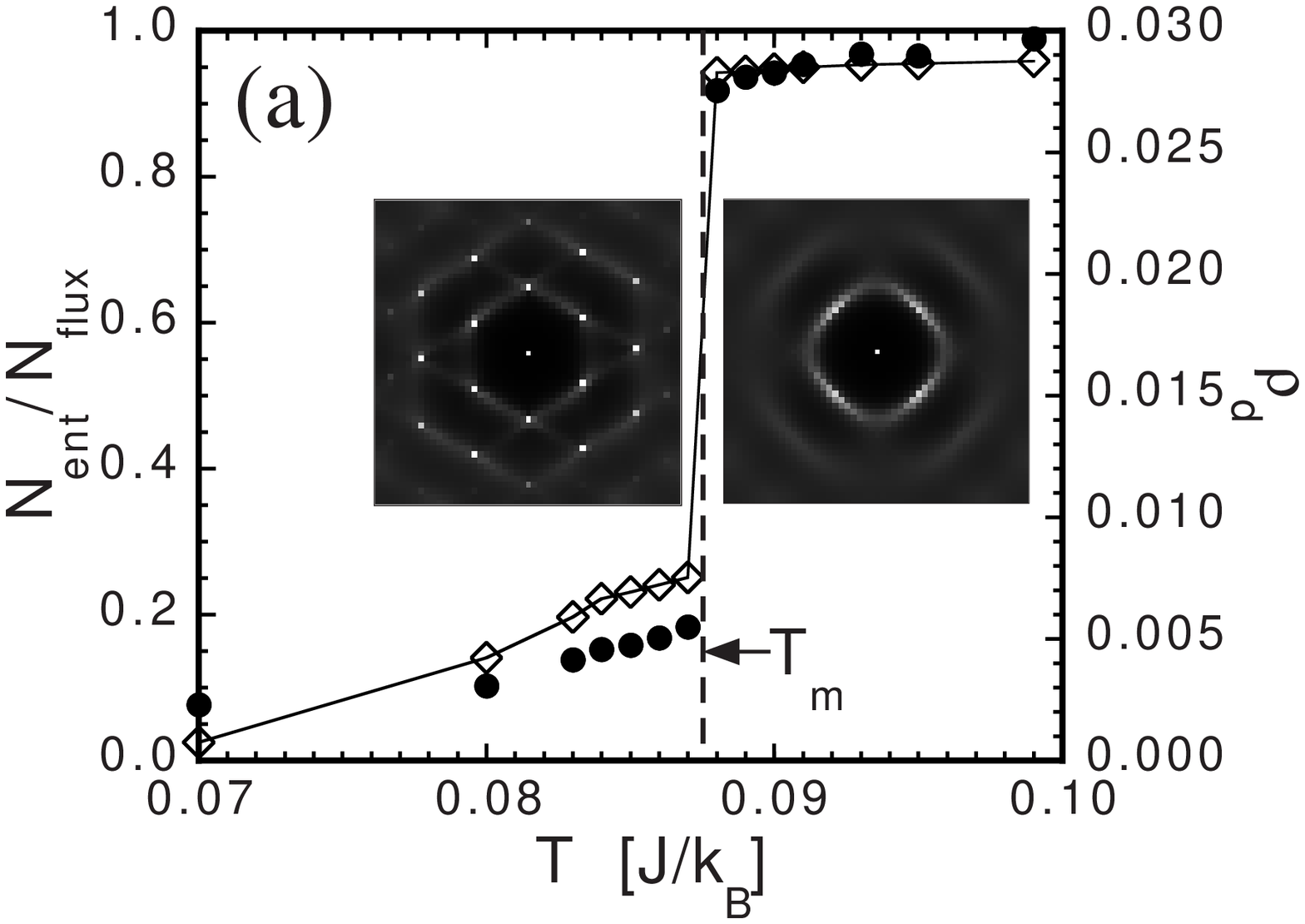}
\epsfxsize = 7.2cm
\epsffile{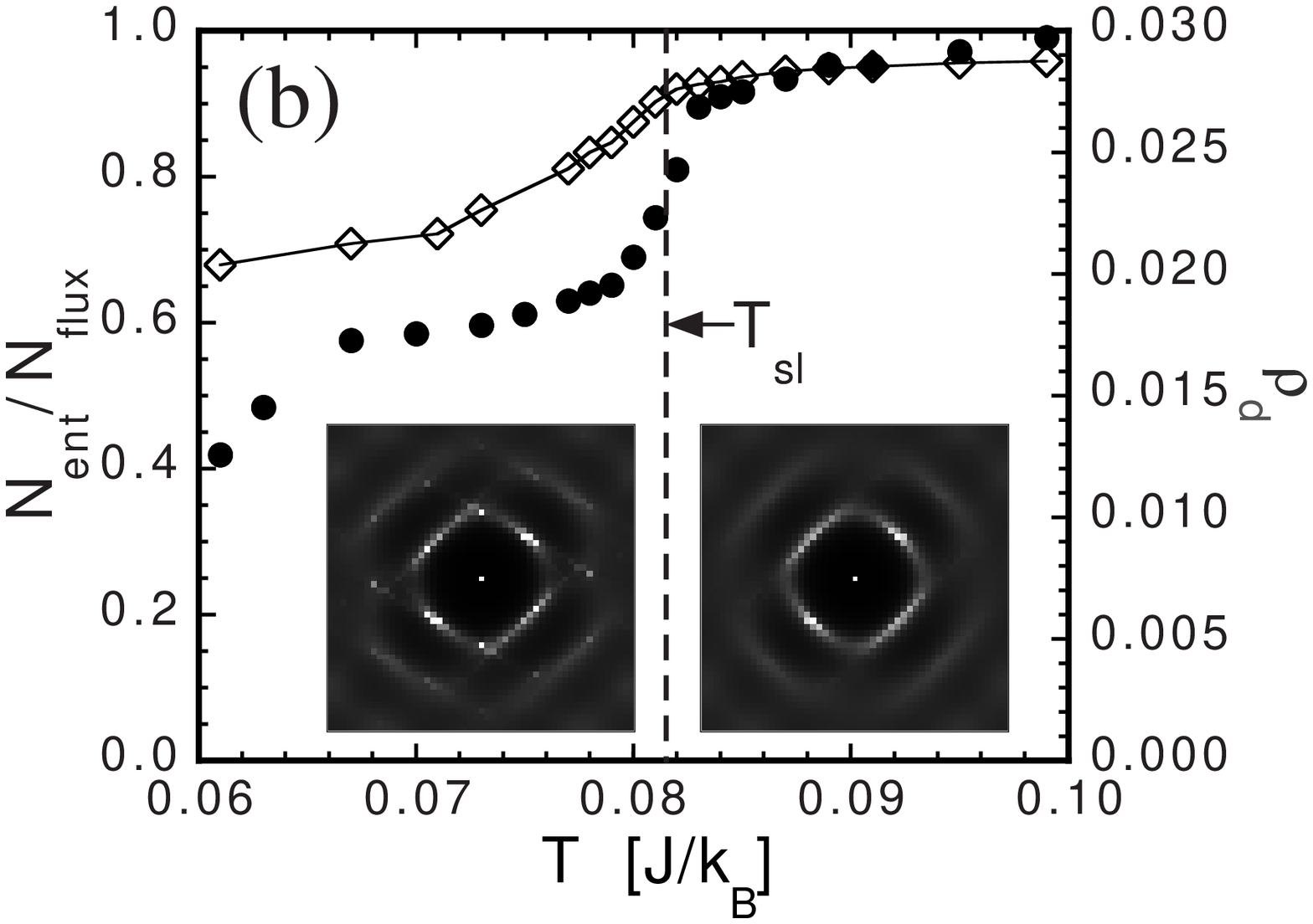}
\vspace{0.2cm}
\caption{
Temperature dependence of $N_{\rm ent}/N_{\rm flux}$ 
($\diamond$) and $\rho_{\rm d}$ ($\bullet$) at 
(a) $\epsilon=0.05$ and (b) $\epsilon=0.11$. 
Structure factors at $T=0.087$ and $0.088 J/k_{\rm B}$ 
are displayed in (a), and those at $T=0.080$ and 
$0.083 J/k_{\rm B}$ are displayed in (b).
}
\label{vorfig}
\end{figure}

Direct observation of flux lines also reveals the 
difference between these two first-order transitions. 
Temperature dependence of $N_{\rm ent}/N_{\rm flux}$ and 
$\rho_{\rm d}$ at $\epsilon=0.05$ and $\epsilon=0.11$ 
is displayed in Figs.\ \ref{vorfig}(a) and \ref{vorfig}(b), 
respectively. $N_{\rm ent}/N_{\rm flux}$ shows a sharp 
jump at $T_{\rm m}$ (Fig.\ \ref{vorfig}(a)) as in pure 
systems, \cite{Nono99} while its temperature dependence 
is continuous around $T_{\rm sl}$ (Fig.\ \ref{vorfig}(b)). 
The quantity $\rho_{\rm d}$ shows sharp jumps both at 
$T_{\rm m}$ and $T_{\rm sl}$ as predicted by Kierfeld and 
Vinokur. \cite{Kierfeld00} These properties are consistent 
with the structure factors shown in the same figures. 
A ring-like pattern is seen in the VL phase as in pure 
systems \cite{Hu} both at $\epsilon=0.05$ and $0.11$. 
The clear triangular Bragg pattern for $T<T_{\rm m}$ at 
$\epsilon=0.05$ represents the formation of a hexatic 
quasi long-range order. The obscure Bragg pattern with a 
$6$-fold symmetry for $T<T_{\rm sl}$ at $\epsilon=0.11$ 
might stand for domains of a short-range hexatic order 
in the $ab$ plane divided by dislocations. 

\begin{figure}
\epsfxsize = 6.5cm
\epsffile{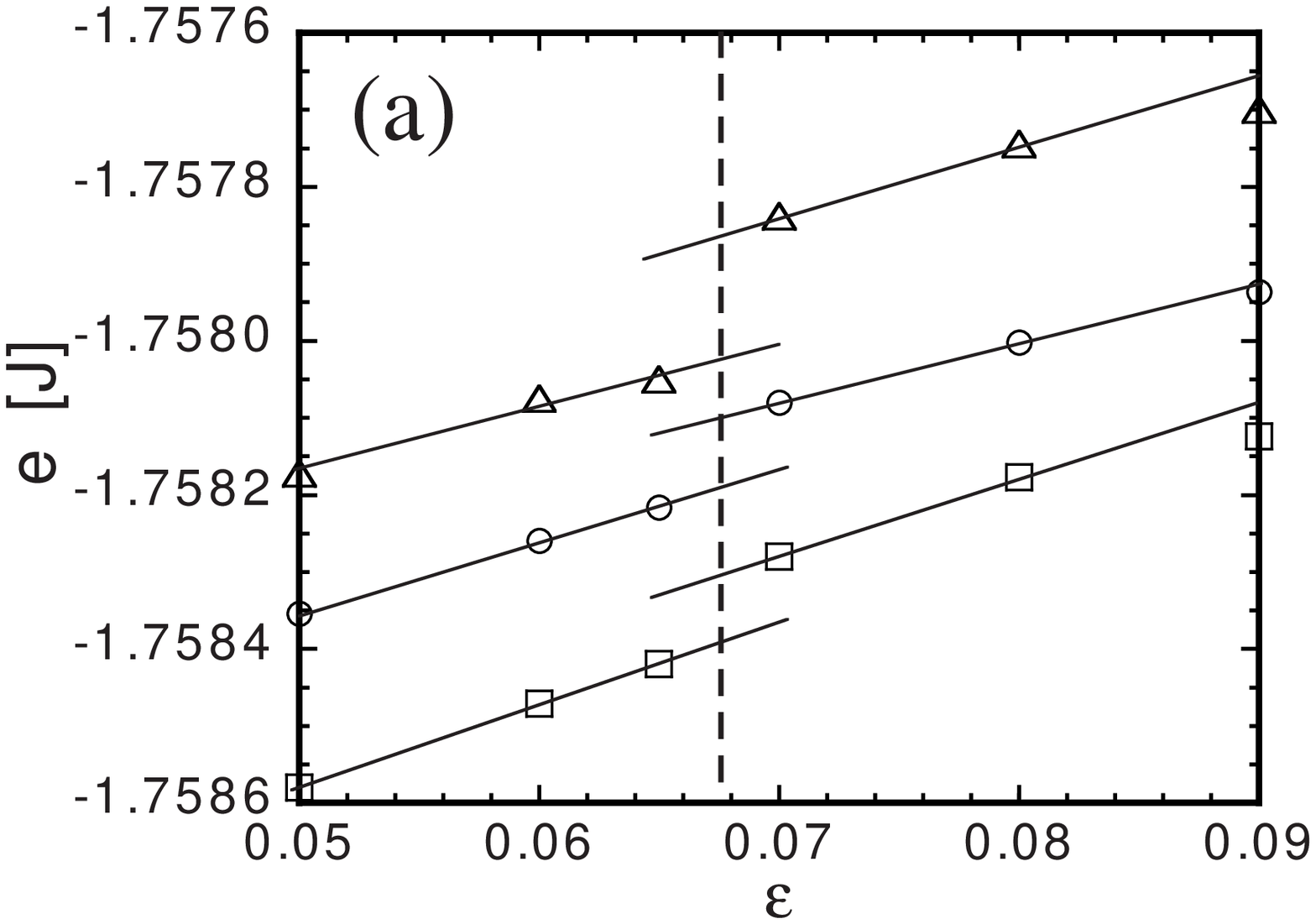}
\epsfxsize = 6.5cm
\epsffile{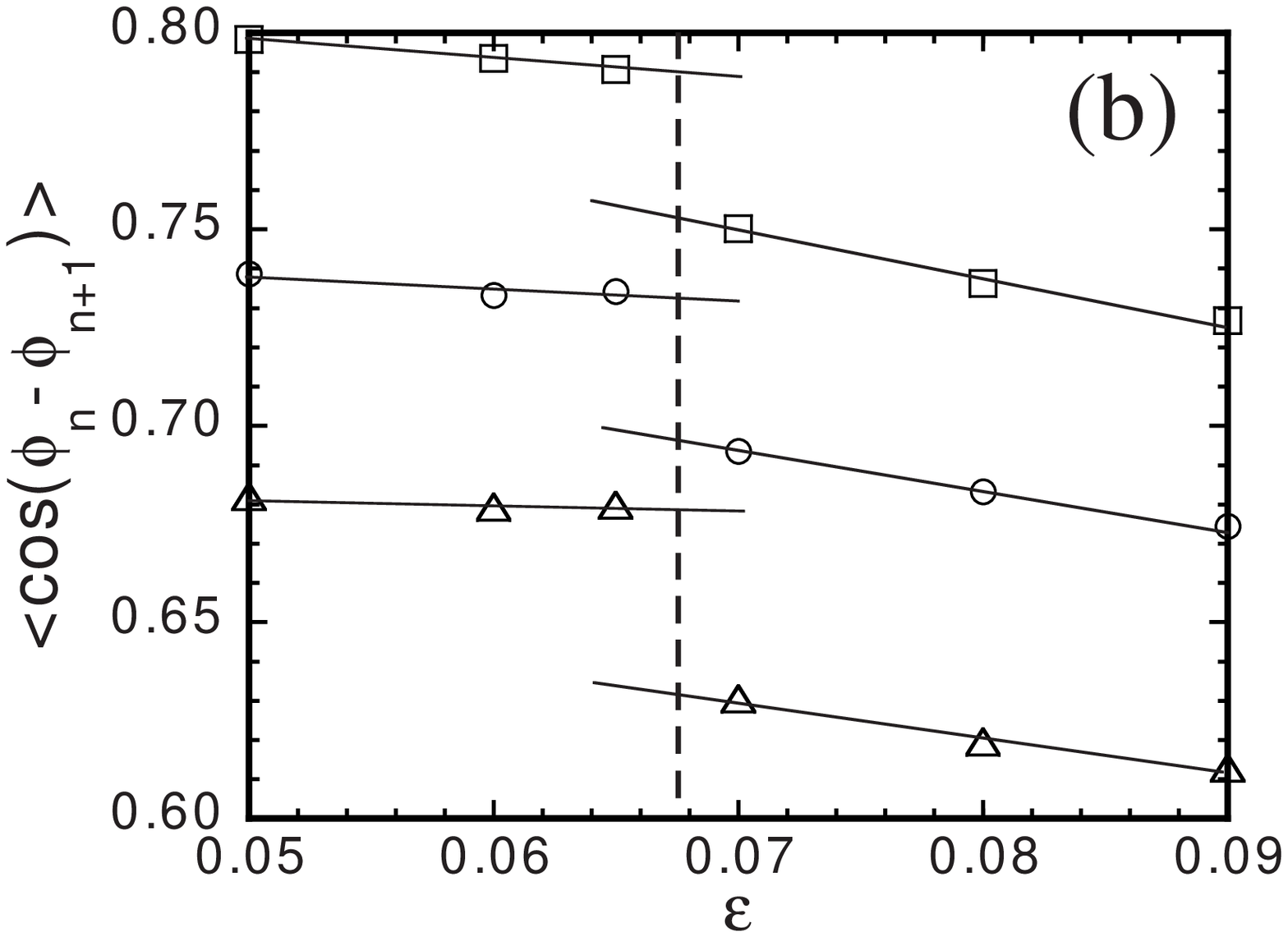}
\vspace{0.2cm}
\caption{
Pinning-strength dependence of (a) $e$ and 
(b) $\langle\cos(\phi_{n}-\phi_{n+1})\rangle$ 
at $T=0.06$ ($\Box$), $0.07$ ($\circ$) and 
$0.08 J/k_{\rm B}$ ($\triangle$). 
The origin of $e$ is varied for each $T$ in (a). 
}
\label{BGVGfig}
\end{figure}

{\it BG--VG and BG--VS phase transitions}.---Pinning-strength 
dependence of $e$ and $\langle\cos(\phi_{n}-\phi_{n+1})\rangle$ 
is shown in Figs.\ \ref{BGVGfig}(a) and \ref{BGVGfig}(b) for 
$T=0.06$, $0.07$ and $0.08 J/k_{\rm B}$. Finite latent heats 
$Q$ indicate that these phase transitions are of first order. 
The quantity $\langle\cos(\phi_{n}-\phi_{n+1})\rangle$ jumps 
simultaneously, which is consistent with Gaifullin {\it et al.}'s 
JPR experiment of BSCCO. \cite{Gaifullin00} Since the anomalies 
are always observed between $\epsilon=0.065$ and $0.070$, 
the phase boundary is almost independent of temperature 
as shown in Fig.\ \ref{phasefig}. We also find a sudden 
jump of $N_{\rm ent}/N_{\rm flux}$ on this phase boundary 
as predicted theoretically. \cite{Ertas96,Giamarchi97} 

The jumps of $e$ and $\langle\cos(\phi_{n}-\phi_{n+1})\rangle$ 
(abbreviated as $\Delta\langle\cos\rangle$ from now on) on 
this almost-flat phase boundary are about one order smaller 
than those on the melting line (see Figs. \ref{VSVLfig}(b) and 
\ref{BGVGfig}(b)). The ratio of the jump of the Josephson energy 
$\Delta e_{\rm J}=-(J/\Gamma^{2}) \Delta\langle\cos\rangle$ to 
$Q$ reveals the difference between these two transitions clearly. 
As in extremely anisotropic pure systems, \cite{Koshelev97} 
this ratio is about one half on the melting line, 
e.g.\ $\Delta e_{\rm J}\approx 5.5\times 10^{-4}J$ and 
$Q\approx 1.1\times 10^{-3}J$ at $\epsilon=0.05$ (see 
Figs.\ \ref{VSVLfig}(a) and \ref{VSVLfig}(b)). On the other hand, 
$\Delta e_{\rm J}/Q$ is about unity on this almost-flat phase boundary, 
e.g.\ $Q \approx \Delta e_{\rm J} \approx 9.0\times 10^{-5}J$ at 
$T=0.07 J/k_{\rm B}$ (see Figs.\ \ref{BGVGfig}(a) and \ref{BGVGfig}(b)). 
The latter result shows a sharp contrast to the JPR experiment of BSCCO, 
where $\Delta e_{\rm J}/Q \gg 1$ seems to be satisfied. \cite{Matsuda01} 

{\it Discussions}.---On the melting line of pure systems, 
$\Delta\langle\cos\rangle$ is proportional to $\Gamma^{2}f$ 
and gradually approaches a saturated value, \cite{Nono00} 
$\Delta\langle\cos\rangle\approx 0.3$. \cite{Shibauchi99,Gaifullin00} 
When the anisotropy is as small as that of YBCO ($\Gamma\approx 7\sim 8$), 
$\Delta\langle\cos\rangle$ is small both on the melting line and 
on the BG--VG phase boundary. In the present system ($\Gamma=20$), 
$\Delta\langle\cos\rangle$ ($\approx 0.22$ at $\epsilon=0.05$) 
is as large as the saturated value on the melting line, while 
it is small on the BG--VG/VS phase boundary. Therefore, a jump 
in $\langle\cos(\phi_{n}-\phi_{n+1})\rangle$ inevitably occurs 
in the VL region, which results in the VS--VL phase transition. 
On the other hand, when the anisotropy is as large as that of 
BSCCO ($\Gamma\geq 150$), $\Delta\langle\cos\rangle$ has reached 
the saturated value both on the melting line and on the BG--VG 
phase boundary as shown experimentally. \cite{Gaifullin00} 
In such a case, it might be difficult to observe a jump 
of $\langle\cos(\phi_{n}-\phi_{n+1})\rangle$ outside of 
the BG phase by the JPR. \cite{Shibauchi99,Matsuda01} 

Finally, the present results summarized in Fig.\ \ref{phasefig} 
are compared with theoretical studies related to the VS phase 
in literature. When Worthington {\it et al.} \cite{Worthington} 
proposed the VS--VL transition line as a reminiscent of the 
melting line in pure systems, the BG phase was out of the scope. 
Ikeda \cite{Ikeda96} derived a phase diagram consisting of the 
VG, VS and VL phases, and argued that the BG phase and the VS 
phase cannot coexist. Quite recently, he modified his argument 
and proposed a possible phase diagram including both the BG 
and VS phases. \cite{Ikeda01} However, he simply assumed 
the existence of the BG phase in this article. 
Kierfeld and Vinokur \cite{Kierfeld00} obtained a phase 
diagram consisting of the BG phase and a first-order transition 
line stretching from the melting line into the VL region with 
a critical point. Although they interpreted this transition 
line as the VG--VL one, it turns out to correspond to 
the VS--VL one as shown in the present study. 
Reichhardt {\it et al.} \cite{Reichhardt} numerically found 
a window-glass-like region with diverging time scales and 
a finite correlation length, which might also correspond 
to the VS phase. We believe that the present simulations 
are the first theoretical derivation of the phase diagram 
consisting of four phases (Fig.\ \ref{phasefig}) based on 
thermodynamic quantities and from a single model. Although 
finite-size analyses of $\Upsilon_{c}$ \cite{Olsson00} have 
not been performed at present, our system size, $L_{c}=40$ for 
$\Gamma=20$ and $f=1/25$, is already very large. According to 
the anisotropy scaling, \cite{Blatter92} this size corresponds to 
$L_{c}\approx 113$ for the parameters in Ref.\ \cite{Olsson00},  
$\Gamma=\sqrt{40}$ and $f=1/20$. 

We would like to thank Y.~Matsuda and T.~Nishizaki for communications, 
and A.~Tanaka for comments. One of us (Y.\ N.) also thanks K.~Kadowaki 
and R.~Ikeda for sending their preprints. Numerical calculations 
were performed on Numerical Materials Simulator (NEC SX-5) 
at National Research Institute for Metals, Japan. 
%

\end{document}